# Rapid-scan acousto-optical delay line with 34 kHz scan rate and 15 attosecond precision

O. Schubert,[1,*] M. Eisele,[1] V. Crozatier,[2] N. Forget,[2] D. Kaplan,[2] and R. Huber[1]

[1] *Department of Physics, University of Regensburg, 93040 Regensburg, Germany*
[2] *Fastlite, 1900 route des crêtes, 06560 Valbonne, France*
*Corresponding author: olaf.schubert@physik.uni-regensburg.de*

An optical fast-scan delay exploiting the near-collinear interaction between a train of ultrashort optical pulses and an acoustic wave propagating in a birefringent crystal is introduced. In combination with a femtosecond Er:fiber laser, the scheme is shown to delay few-fs pulses by up to 6 ps with a precision of 15 as. A resolution of 5 fs is obtained for a single sweep at a repetition rate of 34 kHz. This value can be improved to 39 as for multiple scans at a total rate of 0.3 kHz.

Ultrashort laser pulses are now routinely employed in a rich variety of powerful analytical tools in physics, chemistry and biology. Time-resolved pump-probe studies [1], nonlinear optical frequency mixing [2], optical coherence tomography [3] or broadband and ultrabroadband THz time-domain spectroscopy [4-6] are but a few examples of popular concepts. Temporal or spatial resolution is commonly achieved with the aid of a controllable optical delay line: A laser pulse is shifted in time with respect to a reference and the experiment is repeated for different temporal offsets. The adjustable delay is usually implemented by mounting a retroreflector assembly on a mechanical stage and exploiting the geometric path difference for different positions of the stage.

Fast scan rates are desirable in order to reach sufficient recording speeds for multi-dimensional spectroscopy, to minimize the impact of thermal drifts of complex optical setups, or to monitor rapidly evolving systems, such as live biological tissue. In case of mechanical delay lines the inertia of the mirrors mounted on motor- or piezo-driven stages limits the maximal scan rates to a few tens of Hertz. Fast-scan delays based on loudspeaker diaphragms or rotating mirrors [7,8] have reached higher repetition rates up to tens of kHz, yet with limited precision. A conceptually different route has been taken recently by asynchronous optical sampling (ASOPS) [9,10]. Two mode-locked lasers have been electronically stabilized to slightly mismatched repetition rates, thus causing a shot-to-shot increase of the temporal offset between two pulse trains. For ASOPS the scan window is fixed by the repetition rate of the laser $f_{rep}$ to $1/f_{rep}$, which amounts to typical values between 100 ps and several ns. Active electronic stabilization of the oscillators has been sophisticated to allow for a temporal precision as good as 50 fs [9]. This technique has opened exciting possibilities, e.g., in high-resolution spectroscopy [11].

Nonetheless many experiments call for delay concepts reaching a higher temporal precision and shorter scan ranges. In condensed matter physics, for example, ultrafast electron and lattice dynamics typically occur in the energy range of few meV to eV, corresponding to time scales of the order of few femtoseconds to picoseconds [1,2]. Furthermore pump-probe experiments employing single-cycle [12] or even sub-optical-cycle temporal resolution [4,5,7] are now possible. In this context a precision better than one femtosecond and a scan window of no more than a few picoseconds would be optimal. The ultimate scan rate $f_{scan}$ is given by $f_{scan} = f_{rep} / N$, where N denotes the number of sample points. In this case, each laser pulse creates exactly one data point. Typical values N = 1000 and $f_{rep}$ = 40 MHz suggest an optimal scan rate $f_{scan}$ = 40 kHz.

Here we introduce an acousto-optical fastscan delay capable of scanning a temporal window of 6 ps with a precision of 15 as and a repetition rate of 34 kHz. The novel scanning concept is based on an acousto-optic programmable dispersive filter (AOPDF) [13]. It exploits the propagation of the acoustic signal between successive optical pulses: An acoustic shear wave and an optical pulse train from a femtosecond laser oscillator co-propagate in an anisotropic dielectric. The acoustic wave appears stationary to each optical pulse because the acoustic group velocity $v_{sound}$ is several orders of magnitude lower than the speed of light. As a laser pulse reaches the acoustic wave the polarization of the diffracted electromagnetic wave is switched from the ordinary to the extraordinary axis of the crystal (Fig. 1), by acousto-optic interaction. Consequently the velocity of the diffracted optical pulse is changed from $c/n_o$ to $c/n_e$, where c, $n_o$, and $n_e$ are the speed of light, and the group refractive indices for ordinary and extraordinary polarizations, respectively. Depending on the position $x_i$ of interaction, an optical delay

$$t_i = (n_e - n_o) \times (L - x_i) / c \qquad (1)$$

is introduced between the diffracted and the non-diffracted pulse. L denotes the length of the crystal. During the time interval between successive laser pulses the acoustic wave propagates by a small distance $\Delta x = v_{sound}/f_{rep}$, leading to a variation of the positions $x_1$, $x_2$, $x_3$ etc. of acousto-optic interaction from shot to shot. In this way, the optical delay imposed on each diffracted laser pulse shifts in small increments of

$$\Delta t = (n_e - n_o) \times (v_{sound} / c) \times f_{rep}^{-1} \qquad (2)$$

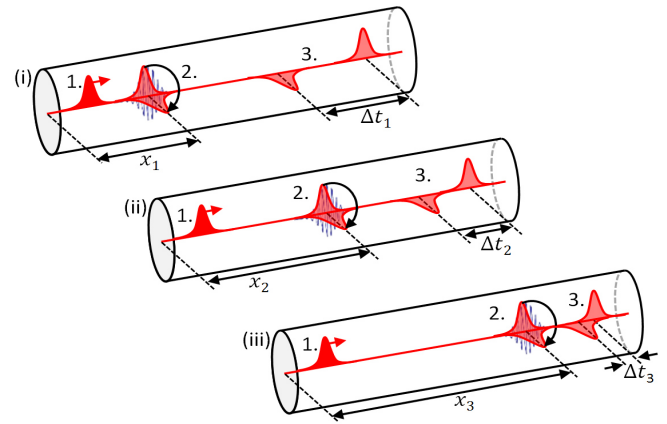

Fig. 1. Principle of operation of the acousto-optical fastscan delay: For laser pulses (red) entering a birefringent crystal (1.), a propagating acoustic wave (blue) appears as a quasi-stationary modulation of the refractive index. Acousto-optic interaction rotates the polarization of the optical pulse by 90°, from the ordinary to the extraordinary axis (2.). While the diffracted pulse proceeds at speed $c/n_e$, the velocity of the non-diffracted pulse amounts to $c/n_o$, leading to a time delay $\Delta t_i$ (3.). Subsequent optical pulses from a high-repetition rate laser system catch up with the propagating acoustic wave at different positions $x_i$, thus imposing different optical delays $\Delta t_i$ ( (i) – (iii) ).

For a standard $TeO_2$ AOPDF and near-infrared wavelengths, $v_{sound}/c = 2.6 \times 10^{-6}$ and $n_e - n_o \leq 0.14$, depending on the cut of the crystal. With $f_{rep} = 40$ MHz, we may choose a delay step $\Delta t$ between 5 fs and 9 fs. In other words, the large velocity mismatch between acoustic and optical waves allows us to convert the laser pulse-to-pulse temporal separation of 25 ns into an optical delay on the femtosecond scale. Note that this acousto-optic device requires neither moving parts nor active electronic feedback loops. Therefore, our concept circumvents problems of beam-pointing fluctuations typically encountered in mechanical scanning systems. Furthermore the time increment $\Delta t$ (Eq. (2)) depends only on material constants and the repetition rate of the laser source, which makes the device intrinsically precise in the attosecond regime, as shown below.

In our implementation, we employ an Er:fiber-laser system operating at a repetition rate of $f_{rep} = 40$ MHz. After the oscillator the pulse train is split into two copies (Fig. 2). One part is directly amplified in a diode-pumped Er:fiber (branch A), while the other one is sent through a $TeO_2$ AOPDF crystal of a length of L = 25 mm (HR25, Fastlite). An acoustic wave is launched by a piezo transducer at a repetition rate of $f_{scan} = 34$ kHz and propagates nearly collinearly with the optical beam. The diffracted femtosecond pulse is coupled into a second Er:fiber stage (branch B). After amplification the pulses in each branch carry energies of more than 8 nJ with pulse durations below 100 fs. For the current crystal length of 25 mm the theoretical maximum of discrete delay positions $t_i$ per scan is $f_{rep}/f_{scan} \approx 1200$. The total delay corresponds to 6 ps, which is scanned with a resolution of 5 fs. These parameters are ideally suited, e.g., for femtosecond pump-probe experiments or electro-optic detection of ultrabroadband THz transients [1,4-6,11].

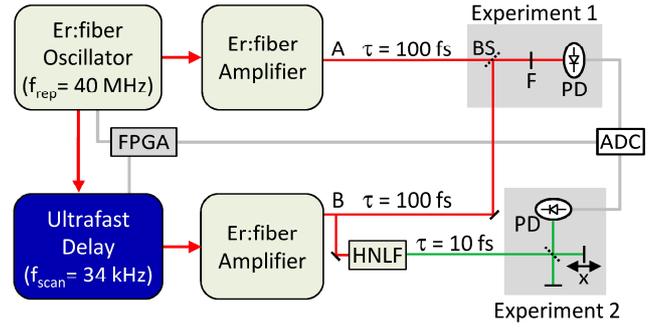

Fig. 2. Acousto-optic fastscan delay incorporated in a femtosecond Er:fiber amplifier system. Seed pulses from an Er:fiber oscillator are split into two branches, one of which is acousto-optically delayed. In experiment 1 we combine pulse trains A and B after amplification with a beamsplitter (BS), filter them spectrally (F) and superpose them on a photodiode (PD). In experiment 2 we characterize pulses compressed in a highly nonlinear fiber (HNLF) via a non-linear autocorrelator, exploiting two-photon absorption and an interferometer with a moveable mirror (x). Electronic synchronization signals are derived directly from the oscillator pulse train and processed by a field programmable gate array (FPGA).

In order to put the precision of the delay to an ultimate test we study interferometric cross-correlation between pulse trains derived from the two amplifier branches. A well defined frequency of the signal is ensured by passing the pulses through a narrowband filter (center wavelength $\lambda_0 = 1560$ nm, full width at half maximum $\Delta\lambda = 12$ nm). Afterwards the collinear superposition of both pulses is detected on an InGaAs photodiode. We digitize the signals with a standard analog-to-digital converter (ADC) and integrate the photodiode signal corresponding to individual laser pulses on a computer. For lowest jitter and noise, a reference clock is obtained from a photodiode monitoring the pulse train from the laser oscillator. Subsequently the clock rate is divided using a field programmable gate array (FPGA) to provide synchronization signals for the ADC and the AOPDF. We subtract the reference signals, obtained by blocking the beam of either branch A or branch B, from the recorded voltage, which yields a bipolar linear cross correlation signal (Fig. 3): Fig. 3a depicts the trace recorded in a single sweep of the delay stage, with a total acquisition time of only 29 µs. Clear indications of interference fringes can be identified. Since the natural increment of the delay line of $\Delta t = 5$ fs is comparable to the oscillation period of the optical carrier wave of 5.2 fs, the optical interference fringes remain undersampled in this scan. The excellent stability and repeatability of the delay, however, allow us to readily increase the effective resolution by means of multiple scans. Following the operating principle of a sampling oscilloscope in electronics, we delay the starting time of the acoustic wave between successive scanning cycles, by an integer fraction of the inverse repetition rate, i.e. by $1/(f_{rep} \times m)$. This translates, via eq. (2), to an additional optical delay $\Delta\tau = \Delta t/m$ for each interleaved scanning cycle with m scans. The number of acquisitions, and thus the effective scan rate and the temporal resolution, can be conveniently selected with a computer interface. A phase-locked loop warrants precise synchronization of the launch event of the acoustic wave and the laser pulse train. The electronic timing jitter of this feedback loop of 30 ps corresponds to an optical jitter of 9 as.

With 128 consecutive scans with an acquisition time of 29 µs, each, we record a cross correlation trace comprising a total number of 148,000 data points within 4 ms. The central part of this signal is shown in Fig. 3b, on a femtosecond scale. Green circles indicate the data obtained with one scan. Gray dots correspond to the signal of individual laser shots during the collection of 128 scans. Featuring an interlaced time increment of $\Delta\tau = 39$ as, the measured data sharply resolve the oscillations due to the near-infrared carrier wave. All data points closely follow a sine function (red curve), demonstrating the extremely high precision as well as the linearity of the delay. This can be seen even better in Fig. 3c, where a close-up of the cross correlation signal is displayed with a total delay window of only 1.3 fs. Even on this scale the data points line up precisely.

To quantify the precision of the delay we fit two cycles of the signal at the center of the cross correlation trace with a sine function (red line in Figs. 3b and 3c). The variance between the measured data points and the fit curve close to the zero crossing is taken as an upper limit of the timing jitter. Besides the actual limitation of the delay line, this value includes mutual timing and intensity fluctuations of the two laser arms during the measurement [14]. The rms value of this jitter is found to be as low as 15 as. This result attests to the excellent stability of the acousto-optic delay and the laser system on time scales up to the acquisition time of 4 ms.

For a test of the long-term timing drift, we repeatedly record linear cross correlation traces for more than 5 hours. For each scan we extract the center of the interference fringes. This analysis yields an rms timing drift of the complete system as low as 0.8 fs (Fig. 3d). No active stabilization, e.g. by additional delay stages or feedback loops, was employed during this measurement. In fact, the long-term jitter is limited by thermal drift of the laser system [14], rather than fluctuations of the acousto-optic delay line itself. If considered necessary, the launch time of the acoustic wave may even be re-programmed for each scan to compensate for laser drift.

Straightforward estimates confirm that the inherent precision of the delay is expected to be even better than 15 as: According to eq. (2), the temporal increment $\Delta t$ is given by the product of a material factor of order $10^{-7}$ and the laser pulse-to-pulse spacing, given by $1/f_{rep}$. During the duration of 29 µs, required for one scan, thermal drifts are negligible and the material parameters are strictly constant. In addition the temperature dependence of the refractive indices $n_o$ and $n_e$ cancels to first order, because only their difference enters. Thus, the pulse-to-pulse timing jitter of the laser system, which typically remains well below 1 fs [6,14], is the dominating factor. From these considerations we estimate the delay precision during one scan to be well below 1 as. Note that we may safely neglect the finite length of the interaction region in eq. (2), because the acousto-optic interaction remains identical for all positions $x_i$. For an interleaved scanning cycle the electronic synchronization becomes the limiting factor, if laser drift can be neglected. As shown above the expected precision in this case is 9 as for our current implementation.

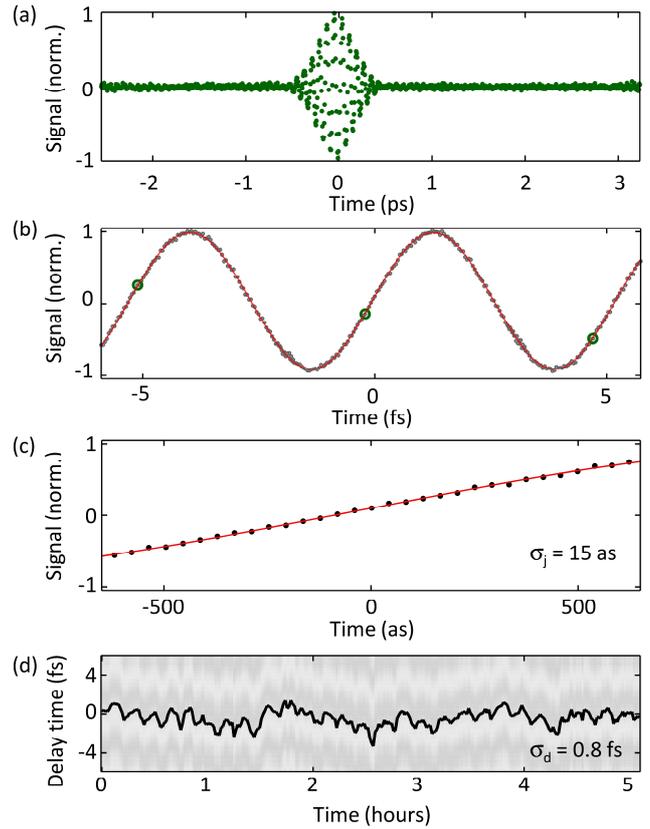

Fig. 3. (a) Linear cross correlation between two spectrally degenerate pulse trains ($\lambda_0 = 1560$ nm, $\Delta\lambda = 12$ nm), as recorded with one laser shot per data point (acquisition time: 29 µs). (b) Close-up with improved time resolution. Green circles indicate the scan from (a), gray dots correspond to data obtained from 128 consecutive, time-offset scans (148,000 data points, acquisition time: 4 ms), red line: fit of sine function to data. (c) Zoom into data from (b), $\sigma_j$ is the rms jitter, (d) Long-term drift of the delay time leading to an rms deviation of $\sigma_d = 0.8$ fs.

One of the outstanding advantages of telecom-compatible Er:fiber femtosecond lasers is their versatility to compress 100-fs optical pulses close to their ultimate single-cycle limit, by means of highly nonlinear germano-silica optical fibers (HNLF) [12,15]. For these ultrabroadband pulses it is particularly important to ensure that the dispersion does not change significantly during a full delay scan. In order to test the compatibility of our fastscan device with such applications, we couple the 100-fs pulses from amplifier B into a dispersion-tailored HNLF assembly (see Fig. 2) to generate a supercontinuum spanning more than one optical octave [12]. A prism sequence is used to compress the short-wavelength part of the spectrum (wavelength range: 950 nm to 1500 nm) close to Fourier-limited pulse duration. Since the spectrum results from a subtle interplay of linear propagation and non-linear effects [15], the duration of the compressed pulses after the HNLF may sensitively depend on the dispersion imposed on the 100-fs input pulses by the fastscan delay.

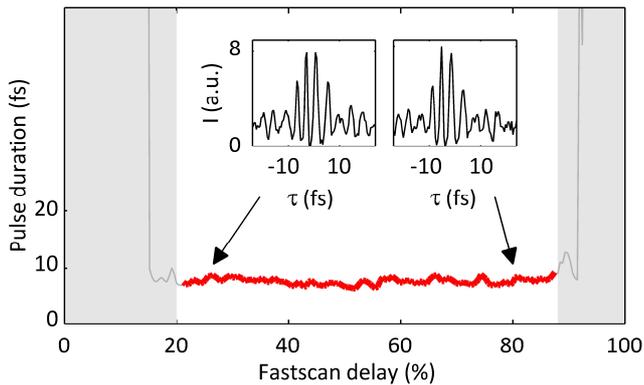

Fig. 4: Compatibility test of the acousto-optic fastscan delay with 8-fs pulses. Duration of pulses derived from a highly nonlinear optical fiber as a function of delay position, as extracted from nonlinear autocorrelation traces assuming a Gaussian pulse shape (red line). The grey area indicates dead time, associated with launching a new acoustic wave. Insets: typical autocorrelation traces for two different delay positions (arrows) of the fastscan delay at the start and the end of the scan, respectively.

We determine the actual duration of the pulses after the prism sequence as a function of the position of the fastscan delay with nonlinear autocorrelation measurements. To this end the pulses are guided into a Michelson interferometer with a standard mechanical delay stage and focused tightly onto a GaAsP photodiode. The two-photon absorption signal is then recorded as a function of the positions of both the fastscan delay and the mechanical delay. For each position of the fastscan delay we extract the approximate pulse duration from the respective autocorrelation trace assuming a Gaussian envelope. The pulse duration of 8 fs is, indeed, constant within the measurement accuracy of ±1 fs, for a duty cycle of about 70 % (Fig. 4). This finding underpins the low variation of the dispersion induced by the delay. The remaining dead-time associated with the launch of a new acoustic wave will be eliminated in a future implementation of the concept.

In conclusion, we demonstrate an ultrastable acousto-optical delay line with 34 kHz scan rate and few-attosecond precision. By translating the repetition rate of the laser into a femtosecond optical delay, our novel concept permits fast and flexible electronic control of sub-optical-cycle delays via precisely adjustable radio frequency signals. Relying on intrinsic material parameters of the acousto-optic cell, such as refractive index and sound velocity, the compact device reaches unsurpassed short- and long-term stability. Since sophisticated acousto-optic pulse shaping has been well established, in a different context [13], we are convinced that the new idea is fully compatible with a broad range of femtosecond lasers, including Er:fiber and Ti:sapphire technology. The novel scheme offers exciting perspectives for a wide field of applications in ultrafast optics, ranging from pump-probe spectroscopy, via THz photonics and OCT to attosecond experiments. More specifically, rapid pump-probe measurements with real-time recording cycles of tens of µs, are anticipated to facilitate, for instance, in situ pump-probe measurements of biological matter. Further work towards extremely sensitive electro-optic sampling of multi-THz transients down to the few photon sensitivity level [16] and multi-dimensional spectroscopy is under way.

Support by the German Research Foundation (DFG) via the Emmy Noether Program (HU1598/1-1) and the European Research Council via ERC Starting Grant QUANTUMsubCYCLE is gratefully acknowledged.

References

1. M. Chergui, A. Taylor, S. Cundiff, R. de Vivie-Riedle and K. Yamagouchi (eds.), Proceedings of the XVIIIth International Conference on Ultrafast Phenomena, EPJ Web of Conferences **41** (2013)
2. S. Mukamel, Annu. Rev. Phys. Chem. **51**, 691 (2000)
3. J. Fujimoto, Nat. Biotech., **21**, 1361 (2003)
4. E. Matsubara, M. Nagai, and M. Ashida, Appl. Phys. Lett. **101**, 011105 (2012)
5. A. Sell, R. Scheu, A. Leitenstorfer, and R. Huber, Appl. Phys. Lett. **93**, 251107 (2008)
6. R. Ulbricht, E. Hendry, J. Shan, T. Heinz, M. Bonn, Rev. Mod. Phys. **83**, 543 (2011)
7. D. Molter, F. Ellrich, T. Weinland, S. George, M. Goiran, F. Keilmann, R. Beigang and J. Léotin, Opt. Exp. **18**, 26163 (2010)
8. N. Chen and Q. Zhu, Opt. Lett. **27**, 607 (2002)
9. R. Gebs, G. Klatt, C. Janke, T. Dekorsy, and A. Bartels, Opt. Exp. **18**, 5974 (2010)
10. S. Kray, F. Spöler, T. Hellerer, and H. Kurz, Opt. Exp., **18**, 9976 (2010)
11. T. Yasui, K. Kawamoto, Y.-D. Hsieh, Y. Sakaguchi, M. Jewariya, H. Inaba, K. Minoshima, F. Hindle, and T. Araki, Opt. Exp. **20**, 15071 (2012)
12. G. Krauss, S. Lohss, T. Hanke, A. Sell, S. Eggert, R. Huber, and A. Leitenstorfer, Nat. Phot. **4**, 33 (2010)
13. P. Tournois, Opt. Comm. **140**, 245 (1996)
14. F. Adler, A. Sell, F. Sotier, R. Huber, and A. Leitenstorfer, Opt. Lett. **32**, 3504 (2007)
15. A. Sell, G. Krauss, R. Scheu, R. Huber, and A. Leitenstorfer, Opt. Exp., **17**, 1070 (2009)
16. G. Günter, A. A. Anappara, J. Hees, A. Sell, G. Biasiol, L. Sorba, S. De Liberato, C. Ciuti, A. Tredicucci, A. Leitenstorfer, and R. Huber, Nature **458**, 178 (2009).